\documentclass[%
  twoside,
  reprint,
  amsmath,amssymb,
  aps,
  pra,
  nofootinbib,
  showpacs,
  superscriptaddress,
  a4paper,
]{revtex4-2}

\usepackage{graphicx}%
\usepackage[usenames,dvipsnames]{xcolor}
\usepackage{siunitx}
\usepackage{booktabs,multirow}
\usepackage{enumitem}
\usepackage{mathtools}
\usepackage{ORCIDinREVTeX}

\usepackage{mathrsfs}
\DeclareMathAlphabet{\mathscrbf}{OMS}{mdugm}{b}{n}
\newcommand{\scrV}{\mathscrbf{V}}

\usepackage{physics}

\usepackage{titlesec}

\usepackage{xmpinclmod}
\includexmp{metadata}

\usepackage[utf8]{inputenc}
\usepackage[T1]{fontenc}

\usepackage{lipsum}

\usepackage[
centering, includefoot,
text={7.1in,10.2in},
total={6.3in,8.75in},
top=0.8in, left=0.62in,
]{geometry}

\usepackage[
  bookmarks=false,
  colorlinks,
  linkcolor=blue,
  urlcolor=blue,
  citecolor=blue,
  plainpages=false,
  pdfpagelabels,
  final,
  breaklinks=true
]{hyperref}
\hypersetup{
pdftitle={Optical polarization skyrmionic fields in free space}, 
pdfauthor={Rodrigo Gutiérrez-Cuevas and Emilio Pisanty}
}

\usepackage{natbib}
\makeatletter \def\NAT@def@citea{\def\@citea{\NAT@separator\,}} \makeatother
\newcommand{\citer}[1]{Ref.~\citealp{#1}}

\newcommand{\reffig}[1]{Fig.~\ref{#1}}

\newcommand{\eqqref}[1]{Eq.~\eqref{#1}}

\newcommand{\bt}{\mathbf}

\newcommand{\bse}{\begin{subequations}}
\newcommand{\ese}{\end{subequations}}
\newcommand{\im}{\text{i}}
\newcommand{\ud}[0]{\mathrm{d}}

\newcommand{\ue}[1]{\hat{\mathbf{e}}_{#1}}

\graphicspath{{figures/},{../figures/}} 
\allowdisplaybreaks

\begin{document}

\title{Optical polarization skyrmionic fields in free space}

\author{Rodrigo Guti\'{e}rrez-Cuevas}
\orcid{0000-0002-3451-6684}
\email{rodrigo.gutierrez-cuevas@fresnel.fr}
\affiliation{Aix Marseille Univ, CNRS, Centrale Marseille, Institut Fresnel, UMR 7249, 13397 Marseille Cedex 20, France}

\author{Emilio Pisanty}
\orcid{0000-0003-0598-8524}
\email{emilio.pisanty@mbi-berlin.de}
\affiliation{ICFO -- Institut de Ciencies Fotoniques, The Barcelona Institute of Science and Technology, 08860 Castelldefels (Barcelona)}
\affiliation{Max Born Institute for Nonlinear Optics and Short Pulse Spectroscopy, Max Born Strasse 2a, D-12489 Berlin, Germany}

\date{March 17, 2021}

\begin{abstract}

We construct optical beams in free space with robust skyrmionic structures in their polarization fields, 
both in the electric spin vector for near-circular fields and in the polarization direction for near-linear fields, 
and for both Bloch (spiral) and Néel (hedgehog) textures.
These structures are made possible by the spin-orbit coupling of tightly-focused nonparaxial optics as applied to higher-order Full-Poincaré beams,
as well as by standing-wave configurations comprising forwards- and backwards-propagating waves.
Our constructions show near-uniform circular and linear polarizations, providing a high degree of topological protection in the absence of nonlinear interactions.
\\[-2mm]

\noindent
\footnotesize
Accepted Manuscript for
\href{%
  https://doi.org/10.1088/2040-8986/abe8b2%
  }{%
  \color[rgb]{0,0,0.55}%
  \textit{J. Opt.} \textbf{23}, 024004 (2021)%
  }, 
available as %
\href{%
  https://arxiv.org/abs/2101.09254%
  }{%
  \color[rgb]{0,0,0.55}%
  arXiv:2101.09254%
  }
under %
\href{%
  %
  https://creativecommons.org/licenses/by-nc-nd/4.0/%
  }{%
  \color[rgb]{0,0,0.55}%
  CC BY-NC-ND%
  }.
\\[-6mm]

\end{abstract}

\maketitle

The field of structured light---the precise shaping of the spatial dependence of the amplitude, phase and polarization of light fields~\cite{gbur2016singularoptics}---has produced, over the past three decades, a wealth of novel structures in the electromagnetic field, from 
phase dislocations~\cite{nye1974dislocations} 
and phase vortices~\cite{allen1992orbital} 
to intricate polarization fields, including 
knots~\cite{kedia2013tying}, 
spirals~\cite{freund2005cones}, 
ribbons~\cite{bauer2019multitwist}, 
M\"obius strips~\cite{freund2010optical1, bauer2015observation, bauer2016optical} and
higher polarization singularities~\cite{pisanty2019knotting}, 
many of which use the spin-orbit coupling of the tight-focusing regime~\cite{bliokh2010angular} to introduce new topological features into the light field. 
Among these structures, one which has sparked interest recently is skyrmions~\cite{tsesses2018optical, du2019deep, gao2019skyrmionic}, and the degree to which they can be recreated within topological photonics.

Skyrmions are topologically-protected excitations that appear in interacting field theories where the field takes values in a sphere;
a skyrmion is then a spatially localized region within which the field completely covers this image sphere.
The simplest version, sometimes called a `baby' skyrmion~\cite{piette1995multisolitons}, is a three-component vector field in two dimensions, such as the magnetization at the surface of a ferromagnet, where a uniform background is punctuated by a region where every field direction is represented.

These excitations were initially described in the context of high-energy physics~\cite{skyrme1962unified, brown2010multifaceted}, where they remain a key connection between QCD and nuclear physics as well as a central tool in string theory and related frameworks.
More recently, they have been the subject of focused interest in magnetic systems~\cite{roessler2006spontaneous, nagaosa2013topological, fert2013skyrmions}---where they hold technological potential in magnetic memories~\cite{fert2013skyrmions} and spintronics~\cite{wiesendanger2016nanoscale}, and have fundamental connections to phenomena like the Hall effect~\cite{jiang2017direct}---%
as well as in cold-atom spinor condensates~\cite{zamora2018skyrmions, ueda2014topological, khawaja2001skyrmion, khawaja2001skyrmions, marzlin2000creation, tuchiya2001topological,leslie2009creation, price2011skyrmion, choi2012observation, choi2012imprinting}, where they form a fertile testing ground for benchmarking topological physics in quantum simulators~\cite{lewenstein2012ultracold}.
More generally, skyrmions have also been observed and used in
liquid crystals~\cite{leonov2014theory}, 
exciton-polariton condensates~\cite{cilibrizzi2016half},
and superconductors~\cite{pershoguba2016skyrmion}, 
and as far afield as neutron stars~\cite{popov2005formation} and other astrophysical phenomena.

Given their importance in multiple domains of topological physics, the generation of skyrmions thus forms a natural target for the toolbox of topological optics.
In this vein, recent experiments have reported 
a skyrmionic lattice in the electric field of a linearly-polarized surface plasmon polariton~\cite{tsesses2018optical},
as well as an isolated skyrmionic singularity in the electromagnetic spin angular momentum vector of an evanescent optical vortex~\cite{du2019deep},
kindling ongoing research into both
linear~\cite{li2020mapping, dai2019ultrafast, shi2020strong}
and spin-based~\cite{tsesses2019spin, davis2020ultrafast, bai2020dynamic}
plasmonic skyrmions.

These structures appear in the evanescent fields near a surface, and they are deeply reliant on the spin-momentum locking of evanescent waves~\cite{bliokh2015quantum, vanmechelen2016universal} to obtain the mixture of transverse and longitudinal vector fields required to form a skyrmionic configuration.
This raises the question, therefore, of whether such structures can appear in propagating fields in free space, in either linear or circular polarizations.
Currently, despite partial negative~\cite{tsesses2018optical} and positive~\cite{du2019deep} answers provided during the initial explorations, this question remains essentially open.

In this work we answer this question in the affirmative, by constructing skyrmionic distributions in the polarization of tightly-focused propagating laser beams in free space:
we present electromagnetic fields containing skyrmionic structures, of both Néel and Bloch types, in the electric spin angular momentum vector of circularly-polarized three-dimensional fields (which we term C-skyr\-mions), as well as in the major axis of fields with polarizations close to linear (which we denote L-skyrmions).

In particular, we present a robust Bloch skyrmion in the electric spin vector of a suitably focused beam, where the polarization field is uniformly near-circularly polarized throughout the focus.
We also show that a C-skyrmion texture of Néel type is also possible, though not as a propagating beam, but relying instead on a standing-wave arrangement 
similar to other localized electromagnetic disturbances constructed recently~\cite{cameron2018monochromatic}.

Moreover, we show that similar standing-wave configurations can produce L-skyrmions with perfectly linear polarization in free space, and without relying on evanescent fields~\cite{tsesses2018optical}.
These L-skyrmions can also be achieved with forward-propagating beams, which causes the polarization to become elliptical; the skyrmionic structures then appear in the major-axis distribution of the field.

We present explicit constructions based on the Complex Focus (CF) fields, \cite{berry1994evanescent, sheppard1998beam, moore2009bases, moore2009closed, gutierrez-cuevas2017scalar, gutierrez-cuevas2018lorentz}, which provide a clean analytical description of a propagating vector beam in the nonparaxial regime.
That said, the robustness of our results indicates that they should be reproducible within Richards-Wolf diffraction theory~\cite{richards1959electromagnetic,novotny2006principles}, and are thus experimentally realizable.

As an important note, the structures that we examine here---like the plasmonic skyrmions presented in earlier work~\cite{tsesses2018optical, du2019deep}---are essentially products of interference between different waves, so they are not `true' skyrmions: there is no physical interaction involving the field which would stabilize the structure and, as such, they are not topologically protected in the formal sense; we thus use the term `skyrmionic field' to mark this distinction.
However, the skyrmionic fields we present here are close to fully circularly- or linearly-polarized over their entire domains, which entails that they are still robust under perturbations, and as close to topologically-protected as possible within their class.
Similarly, this distinction raises the question of whether suitable optical nonlinearities~\cite{kartashov2019frontiers} where the skyrmionic fields we present would propagate as topologically-protected solitons.

In the following, we briefly describe the CF fields we use for our constructions~\cite{CFpackage}, and then address each of the four skyrmionic structures in turn. 
Our implementation is available as \citer{FigureMaker}.

\section{Nonparaxial vector beams}
Skyrmionic fields are fundamentally three-dimensional structures, so they can only be replicated in the polarization structure of a freely-propagating electric field in non-absorbing linear media by considering fully nonparaxial fields.
In vacuum, a general monochromatic field with wavenumber $k$ can be written as a superposition of plane waves travelling in all directions,
\begin{align} 
\label{eq:PWSdecomp}
\bt E (\bt r) 
= 
\int_{4\pi} \bt A ( \bt u) e^{\im k\bt u \cdot \bt r} \ud \Omega,
\end{align}
where $\bt A(\bt u)$ is the plane-wave spectrum (PWS), evaluated at the radial unit vector 
$\bt u = (\cos(\phi) \sin(\theta),\discretionary{}{}{}\sin(\phi) \sin(\theta),\discretionary{}{}{} \cos(\theta))$ 
with spherical polar coordinates $\phi,\theta$; 
to guarantee a solenoidal field with $\nabla \cdot \bt E =0$, the PWS should satisfy the transversality condition $\bt u \cdot \bt A =0 $.

In this work we restrict our attention to skyrmionic structures with cylindrical symmetry about the optical axis, which means that the dependence of $\bt A(\bt u)=\bt A(\theta,\phi)$ on the azimuthal coordinate $\phi$ can be given at most by a global phase, and it is restricted to the form
\begin{subequations}%
\label{eq:genePWS} %
\begin{align} 
\bt A (\bt u)
= & 
\left[ A_p(\theta) \ue{\theta} + A_a(\theta) \ue{\phi}\right] e^{\im m \phi} ,\\
= & 
\left\{ 
  A_p(\theta)
  \left[
    \cos(\theta)(\cos(\phi) \ue{x} +\sin(\phi)  \ue{y}) 
    - \sin(\theta) \ue{z}
  \right]  
\nonumber
\right. \\ & \left. \quad 
  + A_a(\theta)(-\sin(\phi) \ue{x} +\cos(\phi) \ue{y}) 
\right\} 
e^{\im m \phi} 
.
\end{align}%
\end{subequations}%
Here $A_p$ and $A_a$ correspond to the polar and azimuthal components, respectively, of the PWS, and $m\in\mathbb Z$ is the total angular momentum number of the field, which encodes the phase accrued under global rotations about the optical axis.

Under this restriction, the azimuthal part of the Fourier transform in \eqqref{eq:PWSdecomp} can be performed explicitly, yielding the electric fields for the polar and azimuthal parts as
\begin{subequations}
\begin{align}
\!\!\!
\bt E_p(\bt r)
& =
2\pi \im^m
e^{\im m \varphi} 
\!\!
\int_{-1}^1 \!\!\!\! \ud u_z \:
A_p(u_z)
e^{\im u_z kz }
\bigg[
  {-}\im 
  u_z
  \ue\rho
  J_{m}'(u_\rho k\rho)
\nonumber \\ & \qquad 
  + 
  \frac{
    m u_z
    }{
    u_\rho k\rho
    }
  \ue\varphi
  J_{m}(k\rho u_\rho)
  - 
  u_\rho 
  \ue{z}
  J_m(u_\rho k\rho)
  \bigg]
\label{eq:polar-field-final}
\end{align}
and
\begin{align}
\bt E_a(\bt r)
& =
-
2\pi \im^m
e^{\im m \varphi} 
\!\!
\int_{-1}^1 \!\!\!\! \ud u_z \:
A_a(u_z)
e^{\im u_z kz }
\nonumber 
\\ & \quad \times
\bigg[ 
  \frac{
    m
    }{
    k\rho u_\rho
    }
  \ue\rho
  J_{m}(u_\rho k\rho)
  +
  \im 
  \ue\varphi
  J_{m}'(u_\rho k\rho)
  \bigg]
,
\label{eq:azimuthal-field-final}
\end{align}
\end{subequations}
respectively,
in terms of $u_z=\cos(\theta)$ and $u_\rho=\sin(\theta)=\sqrt{1-u_z^2}$, and the cylindrical coordinates $(\rho,\varphi,z)$ of $\bt r$.
Thus,
the azimuthal component of the PWS produces a primarily azimuthal polarization,
whereas the polar component produces a primarily radially and longitudinally polarized field;
since the two parts of the PWS can be set independently, this provides the ability to control the azimuthal component independently of the radial and longitudinal components.

Moreover, the longitudinal component along $\ue{z}$ is caused exclusively by the polar part $A_p(\theta) \ue{\theta}$ of the field, which appears $\SI{90}{\degree}$ out of phase with the radial component.
If $A_p(\theta)$ is real-valued (up to a global phase), then at the focal plane this phase difference carries through to $\bt E(\bt r)$: 
as a general rule, the effect of the longitudinal component is to induce a forwards ellipticity, which is well known as the source of a transverse spin density~\cite{neugebauer2015, bauer2016optical} in nonparaxial optics.

\begin{figure}[b]
\centering
\includegraphics[width=.99\linewidth]{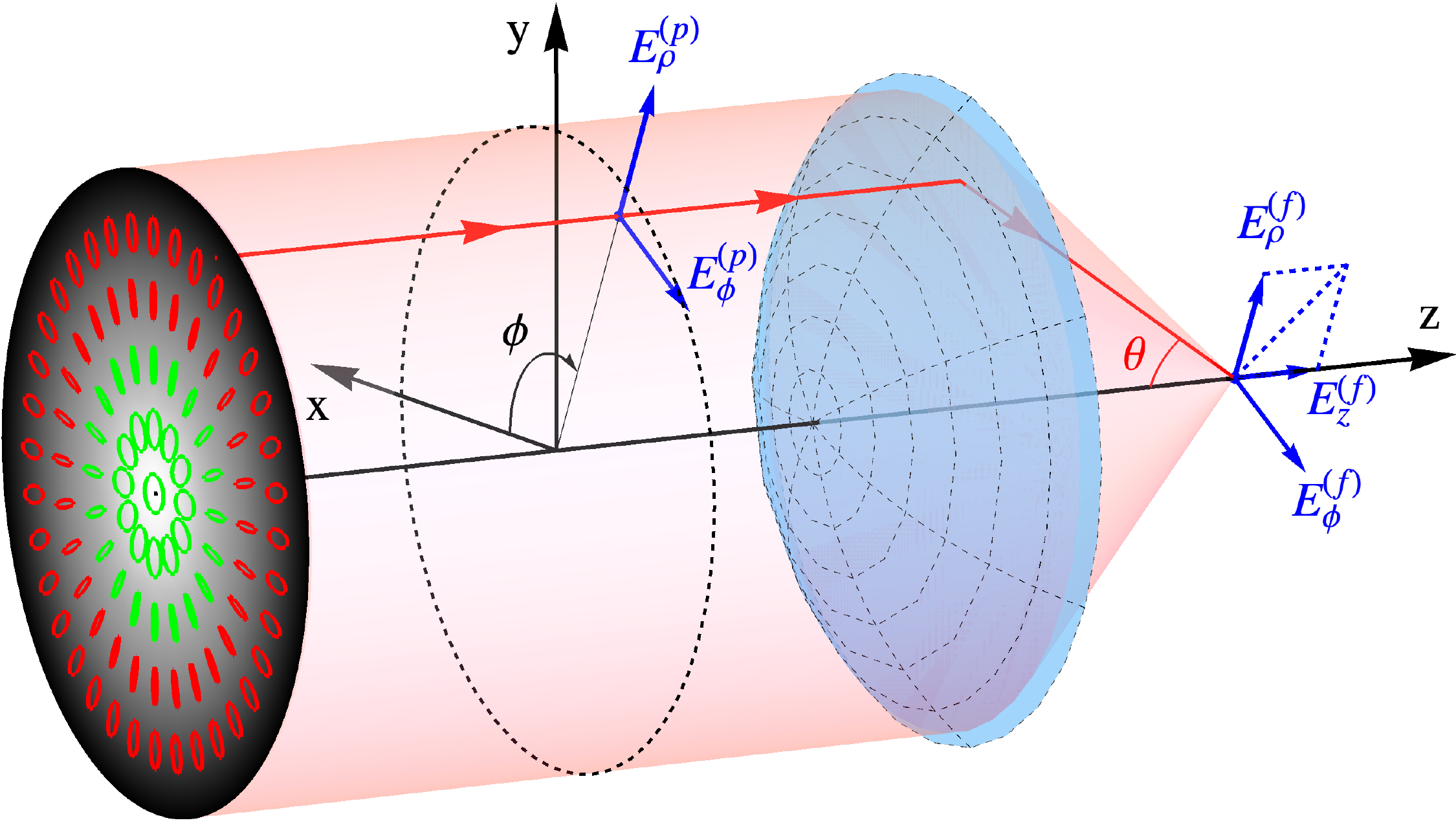}
\caption{
\label{fig:foc} 
Focusing of a paraxial beam by an aplanatic system. To compute the focused field, the incident paraxial beam is decomposed into its radial ($E_{\rho}^{(p)}$) and azimuthal ($E_{\phi}^{(p)}$) components. The radial component is rotated by the lens giving rise to the radial ($E_{\rho}^{(f)}$) and longitudinal ($E_{z}^{(f)}$) components, while the azimuthal component is focused without change. 
}
\end{figure}

These highly-nonparaxial fields are most commonly obtained by focusing paraxial fields using a microscope objective with a high numerical aperture.
This setting is cleanly described using Richards-Wolf diffraction theory~\cite{richards1959electromagnetic, novotny2006principles}, illustrated in \reffig{fig:foc}, which gives $\bt E(\bt r)$ as a (typically numerical) integral over the original paraxial illumination.

A simpler, fully-analytical description, on the other hand, is offered by CF fields~\cite{berry1994evanescent, sheppard1998beam, moore2009bases, moore2009closed, gutierrez-cuevas2017scalar}. 
These are fully-nonparaxial solutions of the (vector) Helmholtz equation, built on the realization that the basic spherical multipolar solutions of the Helmholtz equation,
\begin{align}
\label{eq:multipole}
\Lambda_{l,m} (\bt r) 
= 
4\pi \im^l j_l(kr)Y_{l,m}(\theta, \phi)
\end{align}
are entire functions of the Cartesian coordinates $(x,y,z)$, and they thus remain solutions if one of the coordinates is displaced by a complex offset, to $\Lambda_{l,m}(\bt r - \im \zeta \ue{z})$.
(Here $j_l$ are spherical Bessel functions, $Y_{l,m}(\theta, \phi)$ are spherical harmonics, and $r=\sqrt{x^2+y^2+z^2}$.) 
For small $\zeta$ the solutions are roughly isotropic, and as $\zeta$ increases they take the form of tightly-focused beams, with $\zeta$ corresponding to the Rayleigh range.
In the loosely-focused limit, $k\zeta\gg 1$, $\Lambda_{0,0}$ becomes a paraxial Gaussian beam~\cite{berry1994evanescent, sheppard1998beam, deschamps1971gaussian}, and $\Lambda_{l,\pm l}$ become Laguerre-Gaussian functions without radial nodes~\cite{allen1992orbital, moore2009bases, gutierrez-cuevas2017scalar}.

Finally, these scalar solutions can be elevated to so\-le\-noi\-dal vector solutions by means of the polarization operators
\begin{subequations}%
\begin{align}
\bt V_{\bt p}^{(M)} f(\bt r)
& =
\tfrac{1}{\im k}
\nabla \times (\bt p f(\bt r))
\\
\bt V_{\bt p}^{(E)} f(\bt r)
& =
-\tfrac{1}{k^2}
\nabla \times \left(\nabla \times (\bt p f(\bt r))\right)
\\
\bt V^{(\pm)} f(\bt r)
& =
\bt V_{\ue{\pm}}^{(M)} f(\bt r)
\mp \im \bt V_{\ue{\pm}}^{(E)} f(\bt r)
\end{align}%
\end{subequations}%
of magnetic, electric, and helicity type, respectively, where $\bt p$ is an arbitrary vector and $\ue\pm = \frac{1}{\sqrt{2}}(\ue{x}\pm\im\ue{y})$.
Using these tools we thus obtain
\begin{subequations}%
\begin{align}
\bt E_{l,m,\bt p}^{(M)}(\bt r)
& =
\bt V_{\bt p}^{(M)} \Lambda_{l,m}(\bt r),
\\
\bt E_{l,m,\bt p}^{(E)}(\bt r)
& =
\bt V_{\bt p}^{(E)} \Lambda_{l,m}(\bt r),
\\
\bt E_{l,m}^{(\pm)}(\bt r)
& =
\bt V^{(\pm)} \Lambda_{l,m}(\bt r)
\end{align}%
\end{subequations}%
as the corresponding Complex Focus fields.

In Fourier terms, the scalar multipolar solutions given in \eqqref{eq:multipole} have a PWS given by a pure spherical harmonic $Y_{l,m}(\bt u)$. 
(The vector solutions are obtained, using the 
$\bt u \leftrightarrow \tfrac{1}{\im k}\nabla$
correspondence, by the local polarizations 
$\scrV_{\bt p}^{(M)} = \bt u \times \bt p$,
$\scrV_{\bt p}^{(E)} = \bt u \times (\bt u \times \bt p)$, and
$\scrV^{(\pm)} = \scrV_{\ue{\pm}}^{(M)} \pm\im \scrV_{\ue{\pm}}^{(E)}$
.)
The complex displacement to $\Lambda_{l,m}(\bt r - \im\zeta \ue{z})$ can be introduced directly into \eqqref{eq:PWSdecomp}, where it factors out as the amplitude $e^{k\zeta \: u_z}= e^{k\zeta\cos(\theta)}$.
This factor strongly biases the PWS towards the forwards hemisphere, with a relative power of $e^{-k\zeta}$ remaining in backward-propagating waves, 
similarly to Richards-Wolf configurations, where there is strictly zero power in those modes.

\section{Bloch C-skyrmion}
\label{sec:bloch-c-skyrmion}
\begin{figure}[b]
\centering
\includegraphics[width=0.99\columnwidth]{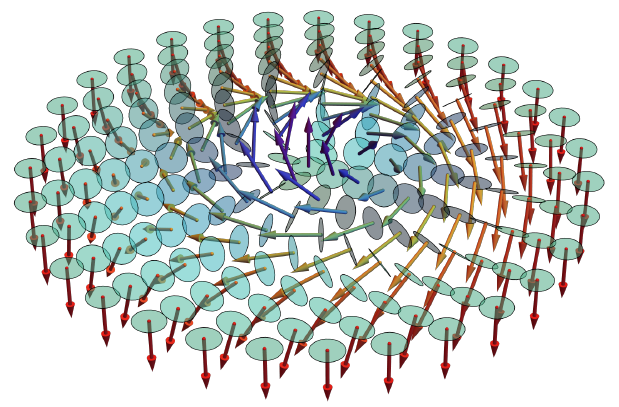}
\caption{
\label{fig:blochcskyrm-ellipses}
Polarization ellipses and the electric-spin distribution for the Bloch C-skyrmion \eqqref{eq:blochcskyrm}.
}
\end{figure}

We now turn to the construction of skyrmionic fields, starting with a Bloch (spiral) skyrmion built from a circularly-polarized field and, specifically, from its electric spin angular momentum, which is orthogonal to the polarization plane.
To make this skyrmion, we require the spin to point forwards at the centre of the beam and backwards at the edge of the focus, connected by a smooth transition during which the spin turns laterally via the azimuthal direction while keeping a zero radial component.
In other words, our beam should be circularly polarized at the centre and at the edges, with opposite helicities, and the polarization ellipse should flip smoothly in between, as shown in \reffig{fig:blochcskyrm-ellipses}.

In particular, this requires a ring of points where the spin is fully azimuthal, corresponding to a circular polarization in the radial-longitudinal plane, which matches the forwards-ellipticity picture for polar fields we just discussed.
When all of this is assembled together, we get the far-field picture shown in \reffig{fig:foc}: a higher-order full-Poincaré beam~\cite{beckley2010full, galvez2012poincare, gutierrez-cuevas2021analytic} with right-circular polarization on the axis, giving way to radial and then left-circular polarizations at larger radii.
This beam is achieved by combining a right-circularly polarized gaussian beam with a left-circularly polarized vortex with orbital angular momentum (OAM) number $\ell=2$.
In the nonparaxial regime, this beam can be written in terms of CF fields as
\begin{align}
\bt E_\mathrm{BCS} (\bt r )
& = 
\cos(\gamma) \:
c_g(\zeta_g)
\bt E_{0,0}^{(+)}(\bt r-\im \zeta_g\ue{z})
\nonumber \\ & \quad \ 
+ 
\sin(\gamma) \:
c_v(\zeta_v)
\bt E_{2,2}^{(-)}(\bt r-\im \zeta_v\ue{z})
,
\label{eq:blochcskyrm}
\end{align}
with independent Rayleigh ranges $\zeta_g$ and $\zeta_v$ for the Gaussian and vortex parts, 
and with a mixing angle $\gamma$ controlling their relative intensity; the normalization constants $c_g(\zeta_g)$ and $c_v(\zeta_v)$ are set to ensure equal power on both components.

The tight focusing preserves the main features of this polarization structure in the transverse plane, and adds a longitudinal component to lift the linear radial polarizations to forward-elliptical, completing the skyrmionic structure.
The full skyrmionic character of this polarization field is most readily evident in the electric spin angular momentum density,
\begin{align}
\bt{S}_\mathrm{E}
= 
\frac{1}{||\bt E||^2}
\text{Im}(\bt E^* \times \bt E)
,
\end{align}
which points in the direction normal to the polarization ellipse, and whose value quantifies the degree of ellipticity, with $0$ corresponding to linear polarization and $1$ to circular polarization.
This electric spin vector is related to the electromagnetic spin density for which evanescent-wave skyrmionic fields have been demonstrated~\cite{du2019deep}, but (as argued previously in the context of superchirality~\cite{vanKruining2018superpositions}) it is more directly relevant for prospective applications that involve dipole interactions with matter.

\begin{figure}[b]
\centering
\includegraphics[width=0.99\columnwidth]{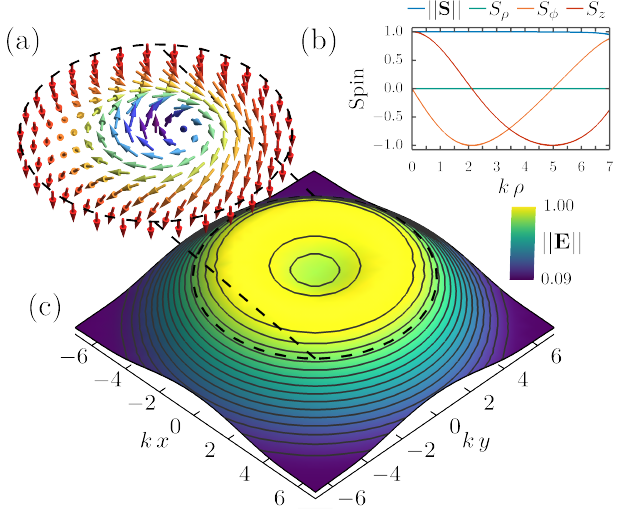}
\caption{
\label{fig:blochcskyrm} 
Bloch C-skyrmion: 
(a) Electric spin distribution, 
(b) spin components and norm, and 
(c) amplitude of the field in \eqqref{eq:blochcskyrm} at the focal plane.
}
\end{figure}

In its basic form, the field structure in \eqqref{eq:blochcskyrm} will produce a nontrivial skyrmionic topology, but it will be far from optimal and present, e.g., points with very low ellipticity.
To address this, we use the free parameters $\zeta_g$, $\zeta_v$ and $\gamma$ to optimize the field, by minimizing the (discretized) integral $\int_0^{\rho_\mathrm{max}} | \bt E(\rho)\cdot \bt E(\rho) |/||\bt E(\rho)||^2 \: \ud\rho$ over the cylindrical radial coordinate $\rho$, to ensure that the field stays as close to circular as possible, under the constrain that $S_z(\rho_\mathrm{max}) < -(1-\epsilon) ||\bt S_\mathrm{E}(\rho_\mathrm{max})||$ to guarantee that the spin changes direction at the edge of the region of interest.

We show this skyrmionic field in Figs.~\ref{fig:blochcskyrm-ellipses} and \ref{fig:blochcskyrm}, 
with the optimized parameters $k \zeta_g\approx 9.20$, $k \zeta_v \approx 4.39$ and $\gamma = \SI{75.39}{\degree}$,
showing 
a roughly flat-top intensity profile (\reffig{fig:blochcskyrm}(c)),
a near-unity electric spin with smooth sinusoidal oscillations in the cylindrical components $S_\rho$ and $S_z$ as well as near-zero~$S_\phi$ (\reffig{fig:blochcskyrm}(b)), and, in particular,
a clear skyrmionic spin distribution (\reffig{fig:blochcskyrm}(a)),
thus establishing a robust Bloch C-skyrmion using tightly-focused propagating waves.

\section{Néel C-skyrmion}
\label{sec:neel-c-skyrmion}
Given this success, it is natural to look for similar constructions where the electric spin of a tightly-focused wave exhibits a Néel (hedgehog) skyrmionic structure, that is, where the spin vector tilts radially instead of azimuthally as it switches from forwards- to backards-facing.
To produce such a structure, we would require the polarization ellipses to have (say) their major axes along the azimuthal direction, and their minor axes along a linear combination of $\ue{z}$ and $\ue\rho$.
In turn, this would require the longitudinal and radial components of $\bt E(\bt r)$ to be strictly in phase with each other (up to a sign).

\begin{figure}[b]
\centering
\includegraphics[width=0.99\columnwidth]{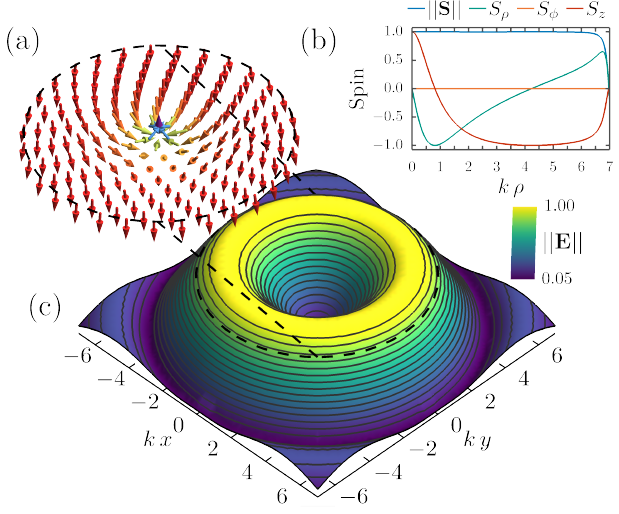}
\caption{
\label{fig:neelcskyrm} 
Néel C-skyrmion: 
(a) Electric spin distribution, 
(b) spin components and norm, and 
(c) amplitude of the field in \eqqref{eq:neelcskyrm} at the focal plane.
The intensity at the origin is 8.85\% of the maximal intensity at $k\rho=3.84$.
}
\end{figure}

However, as mentioned earlier, the natural phase relationship between the radial and longitudinal components of the field is a $\SI{90}{\degree}$ phase delay, as follows from \eqqref{eq:polar-field-final}.
This phase relationship can be altered by choosing a polar amplitude $A_p(\theta)$ with a nontrivial phase in \eqqref{eq:PWSdecomp}, 
or by exploiting the small radial component produced by the azimuthal part $A_a(\theta) \ue{\phi}$ of the field, shown in \eqqref{eq:azimuthal-field-final}, under nonzero total angular momentum number $m\neq0$,
but both of these mechanisms invoke precise cancellations of otherwise fixed quantities, so they can only be expected to work at a discrete set of radii instead of uniformly over the focal plane.

That said, this structure can indeed be achieved, by using a standing-wave configuration and dropping the requirement that the power in the PWS be localized in the forwards direction.
In particular, if the PWS satisfies $A_p(\theta) = A_p(\pi -\theta)$, then the radial component of the field \eqqref{eq:polar-field-final} at the focal plane must vanish;
alternatively, if $A_p(\theta) = -A_p(\pi -\theta)$ then the longitudinal component becomes zero. 
Each of these symmetries thus allows a polar field (with $m=0$) to have a purely longitudinal or purely radial polarization; in combination, they allow us to manipulate the two components independently, and thus to fix the phase relationship between them as required.

To build a concrete example, we use the pure multipolar fields with these symmetries, so 
for the longitudinal component we use the polar field $\bt E_{1,1,\ue{z}}^{(E)}(\bt r)$, with an even $A_p(\theta)$.
For the radial and azimuthal components, we use $\bt E_{1,0,\ue{+}}^{(M)}(\bt r)$ and $\bt E_{3,2,\ue{-}}^{(M)}(\bt r)$, both of which are circularly polarized in the transverse plane, with odd $A_p(\theta)$ and thus no longitudinal component.
Finally, the mixing angles $\gamma$ and $\delta$ in this linear combination,
\begin{align}
\bt E_\mathrm{NCS} (\bt r )
& =  
\cos(\delta)
\left[
\cos(\gamma)
\bt E_{1,0,\ue{+}}^{(M)}(\bt r)
-\sin(\gamma)
\bt E_{3,2,\ue{-}}^{(M)}(\bt r)
\right] 
\nonumber \\ & \qquad
+
\sin(\delta) \:
\bt E_{1,1,\ue{z}}^{(E)}(\bt r)
,
\label{eq:neelcskyrm}
\end{align}
provide the free parameters to control how these components interact to produce the skyrmionic structure.

\reffig{fig:neelcskyrm} shows the resulting field, using mixing angles $\gamma = \SI{87.95}{\degree}$ and $\delta = \SI{7.18}{\degree}$ optimized (as with the Bloch C-skyrmion) to maximize the circularity of the polarization.
This field shows a clear Néel skyrmionic structure in the spin and,
although the intensity at the centre is relatively low,
the spin remains at near-unity throughout the structure.
That said, further improvement of this structure---for instance, using standing-wave Complex Focus fields with appropriate symmetries---is certainly possible.

\section{Bloch L-skyrmion}
\label{sec:bloch-l-skyrmion}
Having constructed skyrmionic structures in the electric spin vector of circularly-polarized fields, it is natural to ask whether the same can be done with linearly-polarized propagating fields,
as has been shown already for evanescent fields produced by plasmonic structures~\cite{tsesses2018optical},
and whether this would require standing-wave or directional configurations.

\begin{figure}[t]
\centering
\includegraphics[width=0.99\columnwidth]{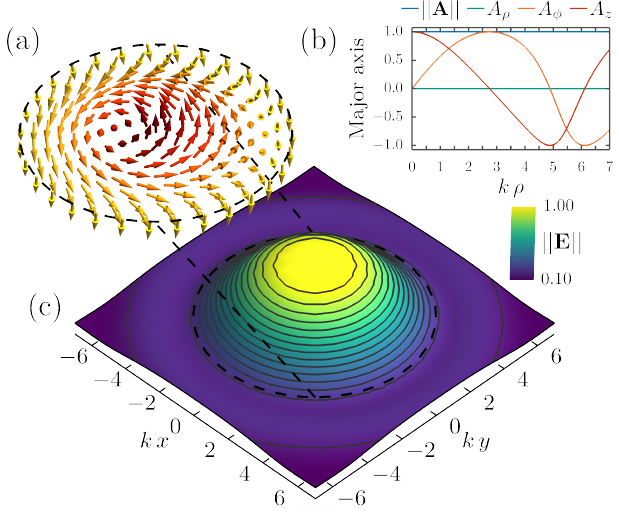}
\caption{
\label{fig:blochlskyrm} 
Bloch L-skyrmion: 
(a) Major axis distribution, 
(b) major axis components and norm, and 
(c) amplitude of the field in \eqqref{eq:blochlskyrm} at the focal plane.
}
\end{figure}

To achieve a Bloch L-skyrmion,  we simply need a field with a longitudinal and azimuthal component that are in phase or $\pi$ out of phase, and this can be easily arranged, since the two components can be fixed independently. 
In general, there will be a radial component coupled to the longitudinal one, with a $\SI{90}{\degree}$ phase difference, but as long as it is smaller than the other two it will only cause the polarization to be elliptical without changing the direction of the major axis. 

This implies that the field will retain some ellipticity, with the skyrmionic structure present in the major axis of the polarization ellipse.
That said, if we optimize the parameters to minimize the spin magnitude (or, alternatively, to maximize the magnitude of the polarization major axis), the resulting field can be made near-linearly polarized.

As a concrete example, we mix the predominantly longitudinal field $\bt E_{0,0,\ue{z}}^{(E)}(\bt r-\im \zeta_p\ue{z})$ together with the azimuthal vector vortex $\bt E_{0,0,\ue{z}}^{(M)}(\bt r-\im \zeta_a\ue{z})$ with a suitable phase:
\begin{align}
\bt E_\mathrm{BLS} (\bt r )
& = 
-
\cos(\gamma) \:
c_p(\zeta_p)
\bt E_{0,0,\ue{z}}^{(E)}(\bt r-\im \zeta_p\ue{z})
\nonumber \\ & \quad \ 
+
\im
\sin(\gamma) \:
c_a(\zeta_a)
\bt E_{0,0,\ue{z}}^{(M)}(\bt r-\im \zeta_a\ue{z})
.
\label{eq:blochlskyrm}
\end{align}
The resulting field, shown in \reffig{fig:blochlskyrm} for the optimized parameters
$k\zeta_p = 0.10$, $k\zeta_a = 1.92$ and $\gamma = \SI{58.50}{\degree}$,
shows a clear spiral skyrmionic structure in the major-axis distribution.

\section{Néel L-skyrmion}
\label{sec:neel-l-skyrmion}
Finally, we turn to the Néel L-skyrmion, a hedgehog structure in the major axis of the polarization field.
In this case, we require a purely polar field in which the relative phase between the longitudinal and radial components of the field is smaller than $\SI{90}{\degree}$, so that the resulting polarization is always elliptical, with a major axis that changes without discontinuities.

\begin{figure}
\centering
\includegraphics[width=0.99\columnwidth]{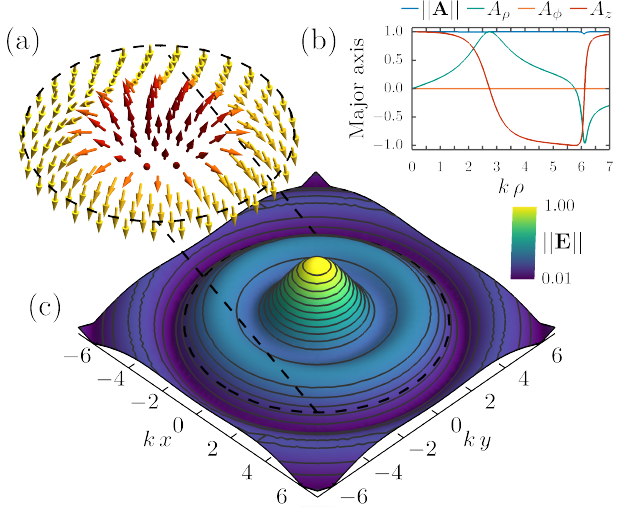}
\caption{
\label{fig:neellskyrm}
Néel L-skyrmion: 
(a) Major axis distribution, 
(b) major axis components and norm, and 
(c) amplitude of the field in \eqqref{eq:neellskyrm} at the focal plane.
}
\end{figure}

To achieve this, we superpose the radial vector vortex $\bt E_{1,0,\ue{z}}^{(E)}(\bt r-\im \zeta_{2}\ue{z})$ together with the purely-longitudinal field $\bt E_{0,0,\ue{z}}^{(E)}(\bt r-\im \zeta_{1}\ue{z})$, allowing their Rayleigh ranges $\zeta_{1}$ and $\zeta_2$, as well as the mixing angle $\gamma$, as free parameters:
\begin{align}
\bt E_\mathrm{NLS} (\bt r )
& = 
\cos(\gamma) \:
c_{1}(\zeta_{1})
\bt E_{0,0,\ue{z}}^{(E)}(\bt r-\im \zeta_{1}\ue{z})
\nonumber \\ & \quad \ 
+
\im
\sin(\gamma) \:
c_{2}(\zeta_{2})
\bt E_{1,0,\ue{z}}^{(E)}(\bt r-\im \zeta_{2}\ue{z})
.
\label{eq:neellskyrm}
\end{align}
The resulting field, shown in \reffig{fig:neellskyrm} for the optimized parameters 
$k\zeta_1 = k\zeta_2 = 0.10$ and $\gamma=\SI{35.85}{\degree}$
(again maximizing the integral of the magnitude of the major axis),
exhibits the required Néel skyrmionic texture,
and has near-perfect linear polarization throughout the region of interest.

In this case, both component fields are weakly directional, with nonzero focusing parameters $\zeta_1$ and $\zeta_2$ but still with significant power in backwards-propagating modes at $\theta>\pi/2$ in the PWS.
This backwards-pro\-pa\-ga\-ting power can be eliminated by increasing the value of $\zeta_1$ and $\zeta_2$: this tends to increase the ellipticity, but it retains the skyrmionic structure in the major axes of the polarization ellipses.

\begin{figure}
\centering
\includegraphics[width=0.99\columnwidth]{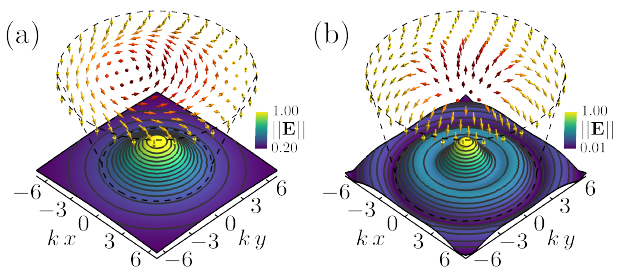}
\caption{
\label{fig:perfectlskyrm}
Major axis distribution and amplitude at the focal plane of the perfect (a) Bloch and (b) Néel L-skyrmions given in Eqs.~(\ref{eq:blochlskyrm},\ref{eq:neellskyrm}) by setting all the $\zeta_j =0$.
}
\end{figure}

Alternatively, it is also possible to make the polarization of the fields exactly linear, by setting all the focusing $\zeta$ parameters to zero, 
which describes a perfect standing-wave configuration with equal power on forwards- and backwards-propagating modes.
We show the resulting skyrmionic fields in \reffig{fig:perfectlskyrm}, for both Bloch and Néel L-skyrmions.

\section{Outlook}
\label{sec:outlook}
The constructions we have presented show that both Bloch and Néel (spiral and hedgehog) skyrmionic textures are possible in both the electric spin vector for quasi-circular fields as well as the major-axis distribution for near-linear polarizations.
These structures are made possible, in part, by standing-wave configurations, while others---most notably the Bloch C-skyrmion---arise in standard propagating beams as a natural consequence of the spin-orbit coupling of nonparaxial optics~\cite{zhu2015transverse}, 
particularly as applied to the tight focusing of higher-order Full Poincaré beams~\cite{yu2020interplay, gutierrez-cuevas2021analytic}, which have also been previously interpreted as forming topological optical skyrmions~\cite{gao2019skyrmionic}.

Moreover, the polarizations in our constructions are almost exactly circular and linear, respectively, and they have high intensities over the polarization singularity, which implies that they are robust with respect to perturbations.
This is important since, in the absence of nonlinear optical interactions giving energy to the topological excitation~\cite{brown2010multifaceted}, optical skyrmions are not `true' skyrmions and thus are not, formally speaking, topologically protected.
Nevertheless, they are still robust, and they are as protected as e.g.\ an optical vortex~\cite{shen2019optical}.

In this regard, it is worth asking whether it is possible to produce perfect C-skyrmions---that is, skyrmionic fields with exactly circular polarization throughout the structure---analogous to the perfect L-skyrmions demonstrated in \reffig{fig:perfectlskyrm}.
The closeness to circularity of our existing constructions, 
together with the topological equivalence of linear and circular polarizations in the Majorana- and Poincarana-sphere representations~\cite{hannay1998majorana, bliokh2019geometric},
indicate that this should be possible.
On the other hand, it is generally harder to achieve perfect circular polarizations, since each component needs to satisfy a specific relation in magnitude as well as in phase, whereas for linear polarizations an appropriate phase relation is sufficient.

In a broader outlook, the optical skyrmions we have presented open the door to the direct creation, detection and manipulation of skyrmions in both magnetic and BEC contexts~\cite{fujita2017ultrafast, hu2015half} using matching structures in light, 
thus allowing for skyrmionic microscopes and optical tweezers~\cite{yang2018photonic, wang2020optical, gutierrez-cuevas2018lorentz},
as well as e.g.\ enabling quantum communication between skyrmionic excitations held in separate cold-atom spinor traps,
or exciting and probing novel resonances in nanoparticle resonators~\cite{zhang2019exploring}.

On a related note, it should also be possible to translate the C-skyrmions we have shown into linearly-polarized structures by using the nonlinear harmonic response or chiral media~\cite{ayuso2019synthetic}, which naturally couples elliptical polarizations with linear polarizations along the ellipse's normal.
This could thus, in principle, directly induce a skyrmionic structure into the field by using an optical source with that structure.
More generally, the possibility of robust optical skyrmionic structures in propagating space opens the possibility for nontrivial effects within nonlinear optics, both in Kerr-type effects as well as in low- and high-order harmonic generation~\cite{watzel2020topological}.

Finally, on the optical side, the presence of Bloch C-skyrmions in tightly-focused beams also raises the possibility that similar structures will appear in suitably thin optical waveguides,
potentially as preserved waveguide eigenmodes, which would further strengthen the potential applications of optical skyrmions for classical and quantum communication.

\section*{Acknowledgements}
We thank Miguel A. Alonso and  Mark R.\ Dennis for helpful discussions.
R.G.-C. acknowledges the Excellence Initiative of Aix-Marseille University - A*MIDEX, a French ``Investissements d'Avenir'' programme.
E.P. acknowledges 
Cellex-ICFO-MPQ Fellowship funding,
and support from 
ERC AdG NOQIA, 
Spanish Ministry of Economy and Competitiveness (%
``Severo Ochoa'' program for Centres of Excellence in R\&D (CEX2019-000910-S), 
Plan National FIDEUA PID2019-106901GB-I00/10.130\allowbreak{}39 / 501100011033, FPI%
), 
Fundació Privada Cellex, 
Fundació Mir-Puig, 
and from Generalitat de Catalunya (AGAUR Grant No.\ 2017 SGR 1341, CERCA program, QuantumCAT \_U16-011424, co-funded by the ERDF Operational Program of Catalonia 2014-2020), 
MINECO-EU QUANTERA MAQS (funded by State Research Agency (AEI) 
PCI2019-111828-2 / 10.13039/\allowbreak{}501100011033), 
EU Horizon 2020 FET-OPEN OPTOLogic (Grant No 899794), 
and the National Science Centre, Poland-Symfonia Grant No.\ 2016/20/W/ST4/00314.

\listoforcidids

\bibliographystyle{arthur} 
\bibliography{references}

\end{document}